\documentclass[prl,aps,amssymb,superscriptaddress,twocolumn]{revtex4-1}
\usepackage{graphicx}
\newcommand{\unit}[1]{\mathrm{~#1}}
\newcommand{\guil}[1]{``#1''}
\newcommand{\moy}[1]{\left < #1 \right >}
\newcommand{\inp}[1]{\left(#1\right)}
\newcommand{\insb}[1]{\left[#1\right]}

\newcommand{\cumul}[2][]{\moy{\moy{{#2}^{#1}}}}
\newcommand{\ioff}[1]{i\inp{#1}}
\newcommand{\icorr}[2]{\moy{\ioff{#1}\ioff{#2}}}
\newcommand{\qicorr}[4]{\moy{\ioff{#1}\ioff{#2}\ioff{#3}\ioff{#4}}}
\newcommand{\intd}{\mathrm{d}}
\newcommand{\der}[3][]{\frac{\intd ^{#1}#2}{\intd #3^{#1}}}
\begin{document}

\title{Noise Intensity-Intensity Correlations and the Fourth Cumulant of Current Fluctuations}

\author{Jean-Charles Forgues}
\affiliation{D\'{e}partement de Physique, Universit\'{e} de Sherbrooke, Sherbrooke, Qu\'{e}bec, Canada, J1K 2R1}
\author{Fatou Bintou Sane}
\affiliation{D\'{e}partement de Physique, Universit\'{e} de Sherbrooke, Sherbrooke, Qu\'{e}bec, Canada, J1K 2R1}

\author{Simon Blanchard}
\affiliation{D\'{e}partement de G\'{e}nie \'{e}lectrique, Universit\'{e} de Sherbrooke, Sherbrooke, Qu\'{e}bec, Canada, J1K 2R1}

\author{Lafe Spietz}
\affiliation{National Institute of Standards and Technology, Boulder, Colorado 80305, USA}

\author{Christian Lupien}
\affiliation{D\'{e}partement de Physique, Universit\'{e} de Sherbrooke, Sherbrooke, Qu\'{e}bec, Canada, J1K 2R1}

\author{Bertrand Reulet}
\affiliation{D\'{e}partement de Physique, Universit\'{e} de Sherbrooke, Sherbrooke, Qu\'{e}bec, Canada, J1K 2R1}

\date{\today}
\begin{abstract}
We report measurements of the correlation between intensities of noise at different frequencies on a tunnel junction under ac excitation, which corresponds to  two-mode intensity correlations in optics. We observe positive correlation, \emph{i.e.} photon bunching, which exist only for certain relations between the excitation frequency and the two detection frequencies, depending on the dc bias of the sample. We demonstrate that these correlations are proportional to the fourth cumulant of current fluctuations.\end{abstract}
\maketitle

Intensity-intensity correlation is a very powerful experimental tool which allows the characterization of statistical properties of light \cite{Loudon}. In particular, it is the quantity that differentiates classical- from quantum light, by showing evidence of photon bunching or anti-bunching. Historically, the first experimental setup which addressed such correlations is that of Hanbury-Brown and Twiss \cite{HBT} on the broad spectrum of a distant star, whose light was measured with two spatially separated detectors. This setup has been repeated many times to characterize light sources (often with narrow spectra) using a beam splitter and again two detectors. Powerful theoretical tools have been developed to understand these measurements, in particular the famous second order correlation function $\moy{I\inp{t}I\inp{t+\tau}}$, which precisely characterizes the correlation between the intensity $I(t)$ of light at time $t$ with that of light at time $t+\tau$. This quantity can be expressed in terms of the creation/annihilation operators of the photon field, taking into account the type of detector that is used (photodetector, homodyne detector, etc.) \cite{Glauber}.

In mesoscopic physics, a lot of effort has been put in the measurement and understanding of current fluctuations \cite{BuBlan,Nazarov_book}. Current or voltage fluctuations are simply another point of view of a fluctuating electromagnetic field, which in the optics community are rather thought of as time dependent electric and magnetic fields. Thus, a noisy electronic device is also a light source, and it is natural to apply tools developed in optics to analyze high frequency electronic noise \cite{Beenakker1,Beenakker2,Blatter}.

In the following we present a measurement of the correlation $G_2=\moy{P_1P_2}-\moy{P_1}\moy{P_2}$ between the powers $P_1$ and $P_2$ of light at two different frequencies in the GHz range, $f_1$ and $f_2$, as proposed in \cite{Blatter}. The source of microwave light is a tunnel junction that is dc biased and excited at frequency $f_0$. We observe that $P_1$ and $P_2$ are correlated when $f_0=f_1+f_2$, $f_2-f_1$ or $(f_2-f_1)/2$, depending on the dc bias voltage of the sample. We demonstrate that $G_2$ is simply proportional to the fourth cumulant of the current fluctuations generated by the tunnel junction. We have calculated this quantity for various excitation voltages and found very good agreement with our data.

\begin{figure}
    \includegraphics[width=0.9\columnwidth] {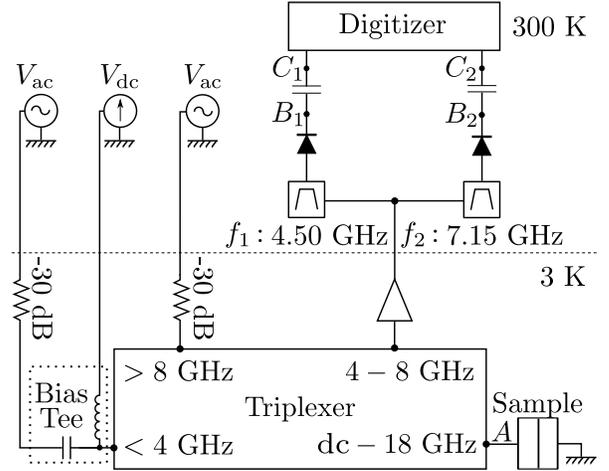}
    \caption{Experimental setup. Diode symbols represent fast power detectors.
    \label{fourthSchem}}
\end{figure}

\emph{Experimental setup.}
We have chosen to perform the measurement on the simplest system that exhibits well understood shot noise, the tunnel junction. The sample is a $R=22\unit{\Omega}$ Al/Al oxide/Al tunnel junction, similar to that used for noise thermometry \cite{Lafe}, cooled at $T=3.0\unit{K}$ so the aluminum remains a normal metal. The experimental setup is depicted on Fig. \ref{fourthSchem}. The combination of a triplexer and a bias tee allows the junction to be connected to different instruments depending on frequency.
The sample is voltage biased by a dc source and connected to two microwave signal generators, one ranging from $10\unit{MHz}$ to $4\unit{GHz}$ and the other above $8\unit{GHz}$. In the $4-8\unit{GHz}$ frequency range, the noise generated by the junction at point $A$ is amplified by a cryogenic amplifier. The setup thus allows the noise to be measured in the $4-8\unit{GHz}$ range while ac excitation can be achieved both below $4\unit{GHz}$ and above $8\unit{GHz}$. This insures that the amplifier never sees the ac excitation of the sample, which avoids spurious signals due to non-linearities in the amplifier at high excitation power.

The amplified noise is split into two branches: in branch 1 (respectively branch 2) the signal is bandpass filtered around $f_1=4.50\unit{GHz}$ with a bandwidth $\Delta f_1=0.72\unit{GHz}$ (resp. $f_2=7.15\unit{GHz}$, $\Delta f_2=0.60\unit{GHz}$). In each branch a fast power detector (diode symbol on Fig. \ref{fourthSchem}) measures the \guil{instantaneous} (averaged over a few ns) noise power integrated in the corresponding bandwidth, \emph{i.e.} $v_{B_k}(t)\propto P_k(t)$ with $k=1,2$. The dc voltage at point $B_k$ is thus proportional to the noise spectral density, $S(f_k)=\moy{|i(f_k)|^{2}}$ with $i(f)$ the Fourier component of the instantaneous current: $\moy{v_{B_k}}\propto S(f_k)\Delta f_k$. A dc block (capacitor on Fig. \ref{fourthSchem}) removes the dc part of $v_B$ to give $v_{C_k}(t)\propto P_k(t)-\moy{P_k} = \delta P_k(t)$. Finally, voltages at points $C_1$ and $C_2$ are simultaneously digitized with a 2-channel, 14-bit, 400 MS/s acquisition card and the correlator $G_2=\moy{\delta P_1 \delta P_2}$ is computed in real time by a 12-core parallel computer, as are the autocorrelations of both channels $\moy{\delta P_k^2}$. The gain has been adjusted to draw on the full dynamic range of the digitizers.

\begin{figure}
  \includegraphics[width=0.9\columnwidth]{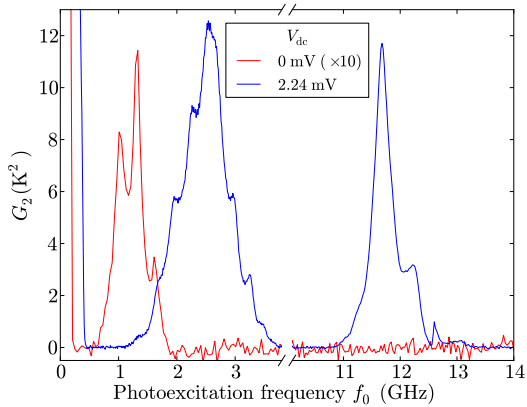}
  \caption{(color online) Reduced $G_2$ vs excitation frequency $f_0$ for two bias voltages: $V_{\mathrm{dc}}=0$ (in red) and $V_{\mathrm{dc}}=2.24$~mV (in blue). The ac excitation amplitude is $V_{\mathrm{ac}}=1.1$~mV for both measurements. The exact shape of the peaks is determined by the frequency response of the experimental setup.
 \label{figfreqSweep}}
\end{figure}

In order to have a quantitative measurement of $G_2$, it is necessary to calibrate the ac excitation voltage reaching the sample and the overall gain of the detection. The calibration of the ac voltage across the sample is performed by measuring the usual photo-assisted noise, \emph{i.e.} $S$ vs $V_{\mathrm{dc}}$, in the presence of a microwave excitation for various excitation voltages \cite{Lesovik,Schoelkopf}. The temperature is large enough ($k_BT\gg hf_{1,2}$) to approximate the noise measured in either branch by its value at zero frequency. In the absence of ac excitation, the noise spectral density is given by $S_0(V_{\mathrm{dc}})= eGV_{\mathrm{dc}}\coth(eV_{\mathrm{dc}}/2k_BT)$ with $G=1/R$, the sample's conductance. Since the excitation frequency $f_0$ is always such that $hf_0\ll k_BT$, the photo-assisted noise can be approximated by its time-average value as if the junction were responding instantaneously to the time-dependent voltage:
\begin{equation}
\bar S(V_{\mathrm{dc}},V_{\mathrm{ac}})=\int_{-\pi}^{\pi} S_0(V_{\mathrm{dc}}+V_{\mathrm{ac}}\cos\theta)\frac{\intd\theta}{2\pi}.
\label{eqSbar}
\end{equation}%
All measurements of $\bar S$ vs $V_{\mathrm{dc}}$ at fixed $V_{\mathrm{ac}}$ are indeed identical and behave according to Eq. (\ref{eqSbar}) whether $f_0=1\unit{kHz}$ or $f_0=12\unit{GHz}$ (data not shown). The setup gain calibration will be discussed later.

\emph{Results.}
The first step for the study of $G_2(V_{\mathrm{dc}},V_{\mathrm{ac}},f_0)$ is to determine the excitation frequency for which $G_2\neq0$. This is done by measuring $G_2$ at fixed bias voltage ($V_{\mathrm{dc}}= 0$ or  $V_{\mathrm{dc}} = 2.4\unit{mV}$) and fixed excitation amplitude ($V_{\mathrm{ac}}= 1.1\unit{mV}$) while varying the excitation frequency $f_0$ from $10\unit{MHz}$ to $3.83\unit{GHz}$ and from $10\unit{GHz}$ to $14\unit{GHz}$. These results are reported in Fig. \ref{figfreqSweep}, where $G_2$ has been reduced to units of $\unit{K^2}$ as noise is often given in terms of equivalent temperature $T_{\mathrm{noise}}=RS/2k_B$. We observe that $G_2$ is large at low frequency for both bias voltages, whereas at higher frequencies $G_2$ strongly depends on $V_{\mathrm{dc}}$. For  $V_{\mathrm{dc}} = 2.4\unit{mV}$, $G_2$ shows peaks at $f_0\simeq f_-=f_2-f_1=2.6\unit{GHz}$ and $f_0\simeq f_+=f_2+f_1=11.6\unit{GHz}$. At zero bias, $G_2$ peaks at $f_0\simeq f_-/2=1.3\unit{GHz}$. Choosing $f_0=f_+/2$ was not possible with this experimental setup since it lies in the $4-8\unit{GHz}$ range.

\begin{figure}
  \includegraphics[width=0.9\columnwidth] {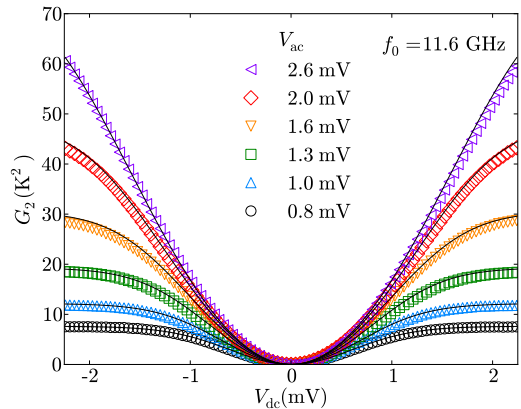}
  \caption{(color online) Reduced $G_2$ vs dc bias voltage for various ac excitation amplitudes at frequency $f_0=f_+=11.6$ GHz. Symbols are experimental data and solid lines theoretical expectations of Eq. (\ref{eqG2}).}
  \label{fig11p6GHz}
\end{figure}

\begin{figure}
  \includegraphics[width=0.9\columnwidth] {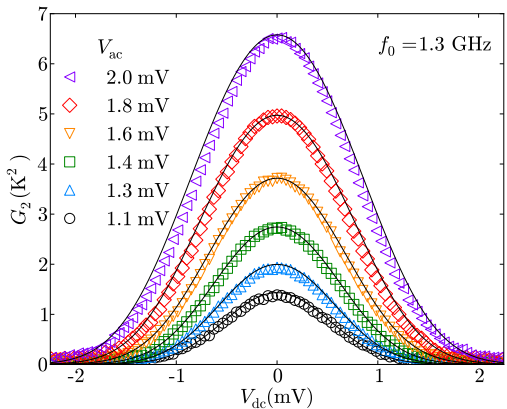}
  \caption{(color online) Reduced $G_2$ vs dc bias voltage for various ac excitation amplitudes at frequency $f_0=f_-/2=1.3$ GHz. Symbols are experimental data and solid lines theoretical expectations of Eq. (\ref{eqG2}).}
  \label{fig1p3GHz}
\end{figure}

We then consider the variation of $G_2$ as a function of the dc voltage for various ac excitation amplitudes at fixed excitation frequencies, chosen according to the position of the peaks we observed. Data on Fig. \ref{fig11p6GHz} correspond to an excitation at $f_0=f_+$. We observe that $G_2$ is maximal at high dc bias and vanishes at $V_{\mathrm{dc}}=0$. We obtained identical results for $f_0=f_-$ (data not shown). Data on Fig. \ref{fig1p3GHz} correspond to $f_0=f_-/2$. Here $G_2$ peaks at 0 dc bias and decays when $|V_{\mathrm{dc}}|$ increases. Moreover, the maximum of $G_2(f_0=f_\pm)$ for a given $V_{\mathrm{ac}}$ is an order of magnitude larger than that of $G_2(f_0=f_-/2)$. $G_2$ measured for low frequency excitation (data not shown) is similar to the behaviour at $f_0=f_\pm$, Fig. \ref{fig11p6GHz}, for small $V_{\mathrm{ac}}$ but differs slightly at higher excitation voltages, as will be discussed below.

\emph{The fourth cumulant of noise.}
In order to describe and understand the meaning of the measured correlator, let us express the noise generated by the sample in Fourier space. Introducing the Fourier component of the power fluctuations $\delta P_k(\varepsilon)$ at frequency $\varepsilon$, one has: $G_2 = \int \intd\varepsilon \moy{\delta P_1(\varepsilon) \delta P_2(-\varepsilon)}$. Here the integral spans over the output bandwidth of the power detectors, $\Delta\varepsilon\simeq100\unit{MHz}$, except for very low frequencies which are removed by the dc block. Specifically, one always has $\varepsilon\neq0$. Power fluctuations are related to current fluctuations by $\delta  P_k(\varepsilon) \propto\int \intd f \insb{i(f)i(\varepsilon - f) + i(-f)i(\varepsilon + f)}$. This integral spans over the bandwidth of the bandpass filters, \emph{i.e.} $f_k\pm\Delta f_k/2$. Thus, $G_2$ is proportional to the correlator between currents at four \emph{different} frequencies which, by definition, is the time averaged fourth \emph{cumulant} of current fluctuations taken at frequencies $f_1$, $f_2$ and $\varepsilon$ (with $\varepsilon\sim0$).
\begin{equation}
 G_2  \propto \qicorr{f_1}{\varepsilon -f_1}{f_2}{-\varepsilon-f_2} \Delta f_1 \Delta f_2\Delta\varepsilon
 \label{eqG2a}
\end{equation}
This represents the fourth cumulant $\moy{\moy{i^4(t)}}=\moy{i^4(t)}-3\moy{i^2}^2$ of the current when it is filtered to keep only frequencies close to $\pm f_1$ or $\pm f_2$ and averaged over time $t$.
In order to detect the fourth cumulant and not the fourth moment of current fluctuations, it is crucial that the four frequencies be different so that correlators $\icorr{f}{-f}$ are never involved. Experimentally, the separation of the signal into two branches with non-overlapping bandwidths insures that $f_1\neq\pm f_2$, while the presence of the dc blocks, by imposing $\varepsilon \neq0$, prevents all other possible occurrences of such correlators.

Single channel autocorrelations $\moy{\delta P_k^2}$ are also related to fourth order current correlators $\qicorr{f}{\varepsilon -f}{f'}{-\varepsilon-f'}$. However, here $f$ and $f'$ belong to the same frequency band, so the correlator is dominated by terms $f=-f'$, and $\moy{\delta P_k^2}\propto \moy{P_k}^2$. The fourth cumulant $G_2$ is only a very small correction to this. Thus, power correlations within the same frequency band are totally dominated by Gaussian fluctuations, as in \cite{M4Portier}. Experimentally, we indeed observe that $\moy{\delta P_k^2}\propto \moy{P_k}^2$, which provides a first way to calibrate our data. Note that since the amplifier noise dominates $\moy{P_k}$, it also dominates $\moy{\delta P_k^2}$ . In contrast, only the fourth cumulant of the amplifier's noise contributes to $G_2$.

\emph{Interpretation.}
We will first focus on the low frequency excitation. When the voltage oscillates very slowly ($f_0<\Delta\varepsilon$), the outputs of the power detectors are proportional to the \guil{instantaneous} noise spectral density $S(t)=S_0\inp{V_{\mathrm{dc}}+V_{\mathrm{ac}}\cos\inp{2\pi f_0t}}$. If $f_0$ is large enough to pass through the dc blocks ($f_0>100\unit{kHz}$), $G_2$ is given by the time average of $\delta S^2(t)$ with $\delta S(t)=S(t)-\bar S$. Here $\bar S$ is the time-averaged value of $S(t)$, \emph{i.e.} the photo-assisted noise of Eq. (\ref{eqSbar}).
In terms of the Fourier components of the current, the amplitude of the oscillation of $\delta S(t)$ at frequency $nf_0$ is given by  $X_n(f,f_0)=\icorr{f}{nf_0-f}$, with $n$ positive or negative integer. This correlator has been calculated and measured in the quantum regime, at very low temperature \cite{GR1,GR2,GR3}. In the high temperature, classical regime that corresponds to the present experiment, $X_n$ reduces to the Fourier coefficient of $\delta S(t)$ oscillating as $\exp (in\theta)$ with $\theta = 2\pi f_0t$, \emph{i.e.}:
\begin{equation}
X_n=\int_{-\pi}^{\pi}S_0(V_{\mathrm{dc}}+V_{\mathrm{ac}}\cos\theta)\exp \inp{in\theta} \frac{d\theta}{2\pi}
\label{eqXn}
\end{equation}%
which is independent of $f$ and $f_0$. The measured fourth cumulant is then given by $G_2\inp{f_0\ll\Delta\varepsilon}=K\sum_n X_n^2$ which, according to the Parseval-Plancherel relation, is the time average of $\delta S(t)^2$. The normalization factor $K\simeq 4\Delta f_1\Delta f_2$ corresponds to the integral over the frequencies that contribute in the setup.
For small $V_{\mathrm{ac}}$, $\delta S(t)\simeq V_{\mathrm{ac}}\der{S_0}{V}\cos(2\pi f_0t)$ oscillates at frequency $f_0$, but for large $V_{\mathrm{ac}}$, $\delta S(t)$  may contain many harmonics of $f_0$. Those harmonics are filtered out as $f_0$ approaches $\Delta\varepsilon$, so $G_2$ should vanish for $f_0\gg\Delta\varepsilon$, as observed in Fig. \ref{figfreqSweep}.
These formulae reproduce our results very well (data not shown) and provide another way to calibrate the overall gain of our setup. This calibration and the one based on $\langle\delta P_k^2\rangle$ are consistent within 10\%, even though they involve integrals over frequencies which slightly differ. In particular, the phase difference between the two branches of the setup, which has been adjusted to maximize the signal, does not influence the autocorrelations $\moy{\delta P_k^2}$ while it  affects the cross-correlation $\moy{\delta P_1\delta P_2}$.

In terms of probability distribution of the current fluctuations one can, at low excitation frequency, think of the amplitude of the current fluctuations as being slowly modulated by the ac voltage: if one considers a Gaussian with a variance oscillating in time, the time-averaged probability distribution is not a Gaussian and has a non-zero fourth cumulant.

To understand why $G_2$ may be nonzero at high excitation frequency, let us first discuss the correlator $\icorr{f}{f'}$. This will be nonzero only for frequencies such that $f+f'=nf_0$ with $n$, an integer. The case $n=0$ corresponds to $\bar S$ whereas $n\neq0$ describes the noise dynamics \cite{GR1,GR2,GR3}. If we consider our experimental setup, relevant frequencies are close to $\pm f_1$ and $\pm f_2$, so this condition becomes $f_1\pm f_2=nf_0$, \emph{i.e.} $f_0=f_\pm/n$.

In such cases, the fourth order correlator of Eq. (\ref{eqG2a}) is dominated by the terms $\icorr{f_1}{\pm f_2}\icorr{-f_1}{\mp f_2}=X_n^2$. Thus, one has:
\begin{equation}
G_2(f_0=f_\pm/n)=K X_n^2
\label{eqG2}
\end{equation}
For small $V_{\mathrm{ac}}$, $X_1^2\approx\frac14\inp{\der{S_0}{V}}^2V_{\mathrm{ac}}^2$ and $X_2^2\approx\inp{\der[2]{S_0}{V}}^2\frac{V_{\mathrm{ac}}^4}{64}$. This is why for $f_0=f_\pm$ we observe on Fig. \ref{fig11p6GHz} that $G_2\simeq KX_1^2$ is 0 at $V_{\mathrm{dc}}=0$ and maximal at large $V$, with $G_2(f_0=f_\pm,V_{\mathrm{dc}}\gg k_BT/e)= K (eGV_{\mathrm{ac}}/2)^2$, whereas for $f_0=f_-/2$, $G_2\simeq KX_2^2$ is maximal at $V_{\mathrm{dc}}=0$, with $G_2(f_0=f_-/2, V_{\mathrm{dc}}=0)=K\inp{e^2GV_{\mathrm{ac}}^2/24k_BT}^2$, and decays at finite dc bias, as seen on Fig. \ref{fig1p3GHz}. We also observed that $G_2(f_0=f_\pm)\propto V_{\mathrm{ac}}^2$ and $G_2(f_0=f_-/2)\propto V_{\mathrm{ac}}^4$ for small $V_{\mathrm{ac}}$ (data not shown). Solid lines on Fig. \ref{fig11p6GHz} and \ref{fig1p3GHz} represent the theoretical predictions of Eqs. (\ref{eqXn}) and (\ref{eqG2}).

\begin{figure}
  \includegraphics[width=0.9\columnwidth] {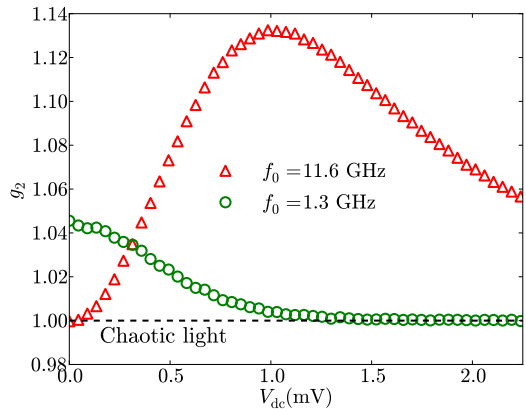}
  \caption{(color online) Measured $g_2$ vs dc bias voltage at $V_{\mathrm{ac}}=1.0\unit{mV}$ for an excitation frequency $f_0=f_+$ (red triangles) or $f_0=f_-/2$ (green circles). The dashed line at $g_2=1$ represents independent photons, \emph{i.e.} chaotic light, while $g_2>1$ corresponds to photon bunching.}
  \label{figxg2}
\end{figure}

In order to make a connection with experiments performed in optics, let us define the  dimensionless correlator:
\begin{equation}
g_2=\frac{\moy{P_1 P_2}} {\moy{P_1}\moy{P_2}}=1+\frac{G_2}{\moy{P_1}\moy{P_2}}
\label{eqxg2}
\end{equation}%
which is usually referred to as the same-time two-mode second order correlator of the electromagnetic field radiated by the junction. To calculate it, one has to subtract the contribution of the amplifiers from $P_k$: $\moy{P_k}\propto(\bar S+S_{amp})$ with $S_{amp}$ the current noise spectral density of the amplifier. The result is plotted on Fig. \ref{figxg2} as a function of $V_{\mathrm{dc}}$ for $V_{\mathrm{ac}}=1.0\unit{mV}$ and both $f_0=f_+$ (red triangles) and $f_0=f_-/2$ (green circles). It is clear from these data that we always observe a positive correlation between the power fluctuations, i.e. photon bunching ($g_2>1$). Each acquisition performed by the digitizer (integrated over $\tau=2.5$~ns) corresponds to an averaged number of $\moy{n_2}=P_2\tau /hf_2\sim21$ photons at 7.2 GHz emitted by the sample at $V_{\mathrm{dc}}=V_{\mathrm{ac}}=1\unit{mV}$ and $f_0=f_+$, plus $\sim40$ photons from the amplifier. Thus our experiment does not measure correlations at the single photon level, but is not very far from that limit, which can be reached by lowering the temperature and the ac power. Note that our power detector, which measures the square of the electric field, cannot differentiate absorption from emission of photons by the sample. This is extremely important when comparing to theories such as \cite{Beenakker1,Beenakker2,Blatter} which consider photon detectors. In particular, the correlations between photon detectors will involve other current correlators \cite{Bednorz}. Whereas $G_2$ at excitation frequency $f_0=f_1+f_2$ probably corresponds to absorption of one photon of frequency $f_0$ and emission of two correlated photons, one at $f_1$  and one at $f_2$, the case of excitation at frequency $f_0=f_2-f_1$ probably corresponds to two photons being absorbed, one at $f_0$ and one at $f_1$ while one photon at frequency $f_2$ is emitted. We indeed observe no difference in $G_2$ for $f_0=f_+$ and $f_0=f_-$.

\emph{Perspectives.}
Current fluctuations generated by a tunnel junction are known to be non-Gaussian, and thus have a non-zero fourth cumulant even in the absence of ac excitation \cite{Levitov}. This should contribute to our measurement as $\cumul[4]{i} \propto e^2S_0$. In the absence of excitation, we indeed observe a non-zero, voltage-dependent $G_2$. However, the signal we observe is much larger than $\langle\langle i^4\rangle\rangle$, and we believe it to be an artefact of our experimental setup (slightly overlapping bandpass filters, cross-talk in the digitizer). This background signal is negligible at large $V_{\mathrm{ac}}$ but has to be subtracted from the measurement at low $V_{\mathrm{ac}}$ for the results to be well fitted by Eqs. (\ref{eqXn}) and (\ref{eqG2}).
Measurements of non-Gaussian noise are also known to be affected by environmental contributions \cite{S3BR,Kindermann,Beenakker}. In \cite{S3BR}, these contributions have been explicitly exhibited by exciting the sample with microwave radiation. Similarly, our results allow us to estimate the environmental contributions to an eventual measurement of the intrinsic fourth cumulant. Each Fourier component of the fluctuating environment will contribute as $\sim X_n^2$. At large voltage, this gives $\sim \langle\delta V^2\rangle e^2G^2\Delta f_1\Delta f_2$ where $\langle\delta V^2\rangle$ is the total (integrated over a potentially huge bandwidth, much wider than the detection one) voltage noise experienced by the sample. It follows that the intrinsic fourth cumulant may be dominated by the environmental contributions.

Another interesting issue raised by our experiment concerns the physical interpretation of $G_2$ in terms of temperature fluctuations. Since the noise can be expressed in terms of a temperature, noise correlations are equivalent to temperature fluctuations which, once integrated over the detection bandwidth of the correlator, have units of $\unit{K^2}$, as in Figs. 2-4. Let us suppose we measure $G_2$ on a macroscopic resistor. Then both $P_1$ and $P_2$ would be proportional to the the same sample's temperature, so $G_2$ would be a measure of the variance of temperature fluctuations, $G_2\propto \langle \delta T^2\rangle$. These are non-zero for any sample of finite size. Thus the thermal noise of a resistor is gaussian only if the sample is infinite, such that its temperature does not fluctuate at all. In other terms, the blackbody radiation should exhibit correlations similar to the one we have observed, which implies that thermal light is not exactly chaotic.

\emph{Conclusion.}
We have shown that the noise power of a photo-excited tunnel junction measured at two different frequencies is substantially correlated. This is crucial in demonstrating the non-gaussian nature of noise signals and a first step towards the study of the intrinsic fourth cumulant of current fluctuations. In terms of photons, we have shown that photo-assisted shot noise corresponds to photon bunching.

We acknowledge fruitful discussions with A. Bednorz, W. Belzig, M. Devoret and J. Gabelli. This work was supported by the Canada Excellence Research Chairs program, the NSERC, the MDEIE, the FRQNT, the INTRIQ and the Canada Foundation for Innovation.

\end{document}